\documentclass[prl,aps,floats,amssymb,twocolumn,preprintnumbers]{revtex4}

\usepackage{amssymb}
\usepackage{epsfig}

\newcommand{\be}{\begin{equation}}
\newcommand{\ee}{\end{equation}}

\newcommand{\chiPT}{$\chi$PT }
\newcommand{\FRRchiPT}{FRR$\chi$PT }

\begin{document}

\preprint{
\vbox{
\hbox{ADP-04-13/T595}
\hbox{JLAB-THY-04-8}
\hbox{DESY 04-075}
}}

\title{Precise determination of the strangeness magnetic moment of the
nucleon}

\author{D.~B.~Leinweber}
\author{S.~Boinepalli}
\author{I.~C.~Cloet}
\author{A.~W.~Thomas$^\dagger$}
\author{A.~G.~Williams}
\author{R.~D.~Young}
\author{J.~M.~Zanotti$^*$}
\author{J.~B.~Zhang}
\affiliation{Special Research Centre for the
Subatomic Structure of Matter,
and Department of Physics,
University of Adelaide, Adelaide SA 5005, Australia}
\affiliation{$^\dagger$ Jefferson Laboratory, 12000 Jefferson Ave.,
Newport News, VA 23606 USA}
\affiliation{$^*$John von Neumann-Institut f\"ur Computing  NIC, \\
Deutsches Elektronen-Synchrotron DESY, D-15738 Zeuthen, Germany}
\begin{abstract}
By combining the constraints of charge symmetry with new chiral
extrapolation techniques and recent low mass lattice QCD simulations
of the individual quark contributions to the magnetic moments of the
nucleon octet, we obtain a precise determination of the strange
magnetic moment of the proton. The result, namely $G_M^s = -0.046 \pm
0.019\ \mu_N$, is consistent with the latest experimental measurements
but an order of magnitude more precise.  This poses a tremendous
challenge for future experiments.

\end{abstract}

\pacs{12.39.Fe, 12.38.Gc, 13.40.Em, 14.20.Dh, 14.20.Jn}

\maketitle


There is currently enormous interest in the determination of the
strangeness content of the nucleon. It is crucial to our
understanding of QCD to determine precisely the role played by
heavier, non-valence flavors. On the experimental side new results on
strangeness in the nucleon have been reported recently from JLab
(HAPPEX) \cite{Aniol:2000at} and MIT-Bates (SAMPLE)
\cite{Spayde:2003nr}. In the near future we can expect even more
precise results from the A4 experiment at Mainz as well as G0 and
HAPPEX2 at JLab. By contrast, the theoretical situation is 
somewhat confused, with the predictions of various quark models covering 
an enormous range. Direct calculations within lattice QCD have not yet 
helped to clarify the situation, with values for $G_M^s$ ranging from 
$-0.28 \pm 0.10$ \cite{Mathur:2000cf} to $+0.05 \pm 0.06$ \cite{Lewis:2002ix}. 

We take a different approach, exploiting the advances in lattice QCD
which have enabled quenched QCD (QQCD) simulations of magnetic moments
at pion masses as low as 0.3--0.4
GeV~\cite{Zanotti:2001yb,Leinweber:2002bw,FLICscaling,FLIClqm}, as
well as the development of new chiral extrapolation
techniques~\cite{Young:2002ib,Young:2004tb}.  Using these techniques we
determine, in full QCD,
the ratios of the valence $u$-quark contribution to the magnetic
moment of the physical proton to that in the $\Sigma^+$ and of the
valence $u$ quark in the physical neutron to that in the $\Xi^0$.
From these ratios, the experimental values of the octet moments and
charge symmetry we deduce a new theoretical value for $G_M^s$ which is
precise -- setting a tremendous challenge for the next
generation of parity violation experiments.
 
\begin{figure}[tbp]
{\includegraphics[height=3.3cm,angle=90]{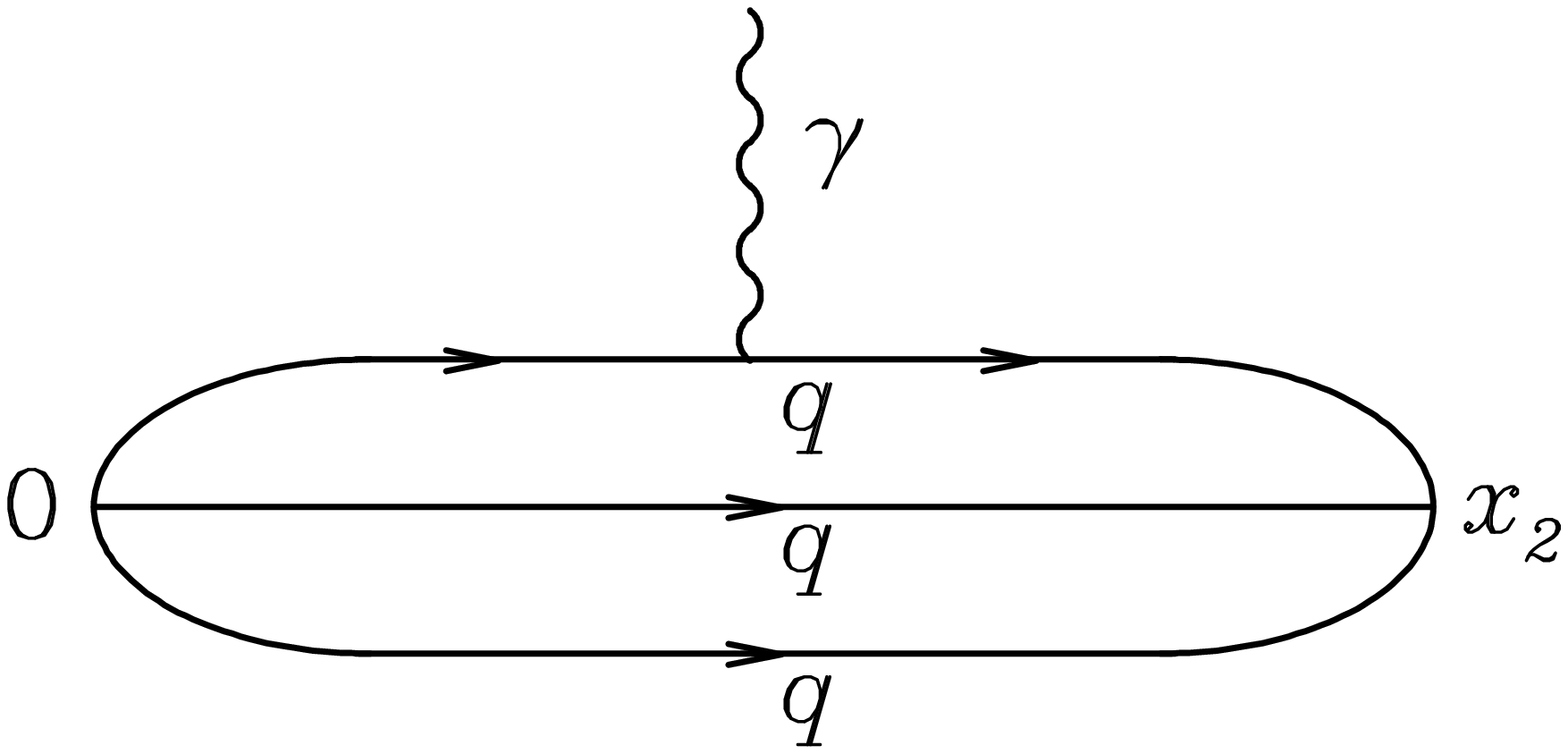} \hspace{0.8cm}
 \includegraphics[height=3.3cm,angle=90]{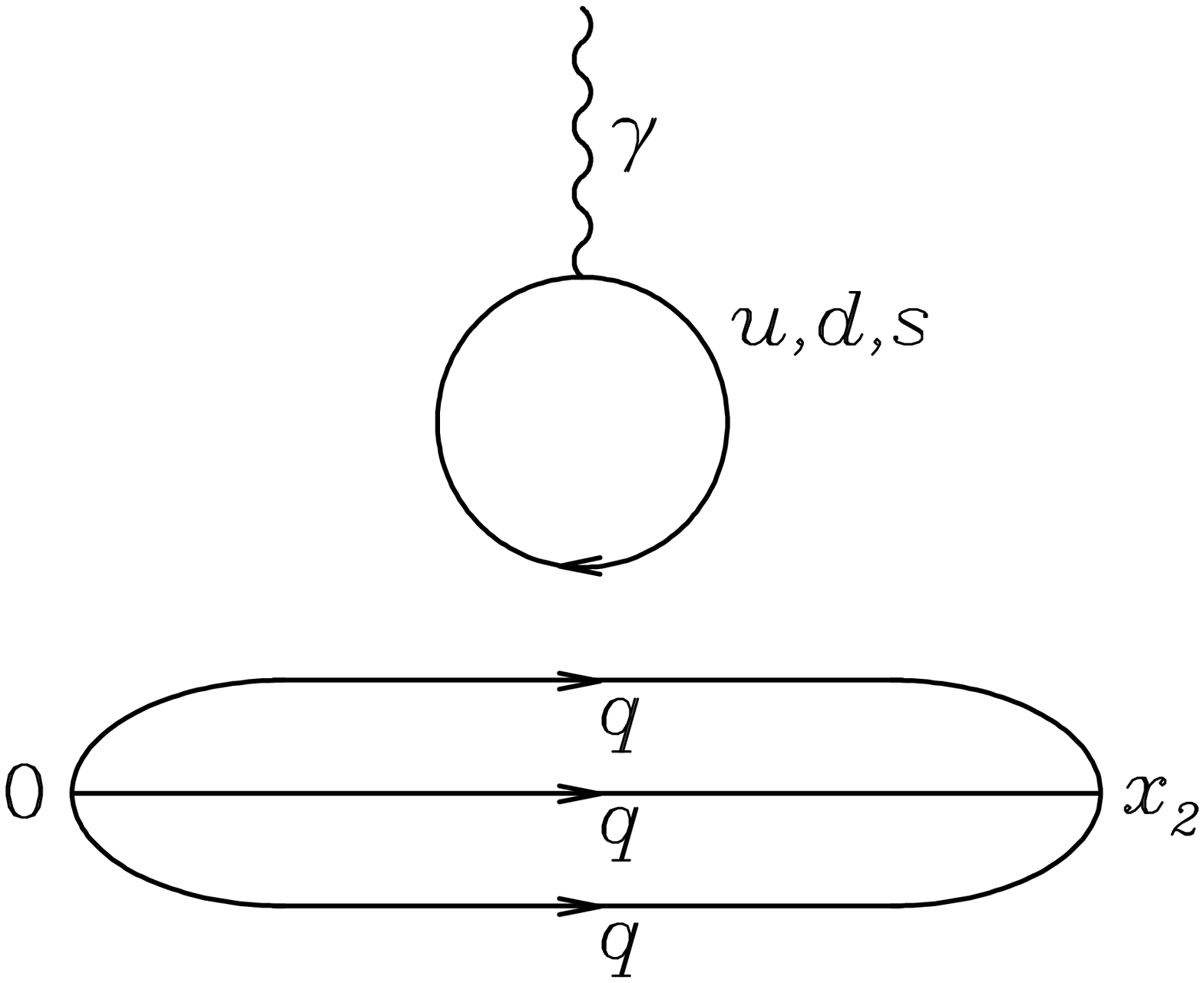}}
\caption{Diagrams illustrating the two topologically different
insertions of the current within the framework of lattice QCD.  
}
\label{topology}
\vspace{-0.5cm}
\end{figure}

As illustrated in Fig.~\ref{topology}, the three point function required 
to extract a magnetic moment in lattice QCD involves two topologically 
distinct processes. (Of course, in full QCD these diagrams incorporate 
an arbitrary number of gluons and quark loops.) The left-hand diagram 
illustrates the connected
insertion of the current to one of the ``valence'' quarks of the
baryon.  In the right-hand diagram the external field couples to a 
quark loop. The latter process, where the loop involves an 
$s$-quark, is entirely responsible for $G_M^s$. 
%

Under the assumption of charge symmetry~\cite{ChargeSymm}, 
the magnetic moments of the octet baryons
satisfy~\cite{Leinweber:1996ie}:
\begin{eqnarray}
p &=& e_u\, u^p + e_d\, d^p + O_N  \, \, ; \, \, 
n = e_d\, u^p + e_u\, d^p + O_N  \, , \nonumber \\
\Sigma^+ &=& e_u\, u^{\Sigma} + e_s\, s^\Sigma + O_\Sigma  \, \, ; \, \, 
\Sigma^- = e_d\, u^{\Sigma} + e_s\, s^\Sigma + O_\Sigma  \, , \nonumber  \\
\Xi^0 &=& e_s\, s^\Xi + e_u\, u^{\Xi} + O_\Xi  \, \, ; \, \,
\Xi^- = e_s\, s^\Xi + e_d\, u^{\Xi} + O_\Xi  \, . \nonumber  \\
\label{equalities}
\end{eqnarray}
Here, $p$ and $\Xi^-$ are the physical magnetic moments 
of the proton and $\Xi^-$,
and similarly for the other baryons. The valence $u$-quark sector
magnetic moment  
in the proton, corresponding to the LHS of Fig.~\ref{topology}, 
is denoted $u^p$. Charge symmetry has been used to replace the
$d$-quark contribution in the neutron by $u^p$, $d$ in the $\Sigma^-$
by $u$ in the $\Sigma^+$ ( $u^\Sigma$), and so on.  
The labels on quark magnetic moments allow for the environment sensitivity
implicit in the three-point function \cite{Leinweber:1996ie,Leinweber:1999nf}.
That is, the naive expectations of the constituent quark model, namely  
$u^p/u^{\Sigma} = u^n/u^{\Xi} = 1$, may not be satisfied.
The total contribution from quark-loops, $O_N$, 
contains sea-quark-loop contributions (RHS of Fig.~\ref{topology}) from
$u$, $d$ and $s$ quarks. By definition
\begin{eqnarray}
O_N &=& \frac{2}{3} \,{}^{\ell}G_M^u - \frac{1}{3} \,{}^{\ell}G_M^d -
\frac{1}{3} \,{}^{\ell}G_M^s \, , \\
&=& \frac{{}^{\ell}G_M^s}{3} \left ( \frac{1 -
{}^{\ell}R_d^s}{{}^{\ell}R_d^s } \right ) \, , 
\label{OGMs}
\end{eqnarray}
where the ratio of $s$- and $d$-quark {\it loops}, 
${}^{\ell}R_d^s \equiv {{}^{\ell}G_M^s}/{{}^{\ell}G_M^d}$, is
expected to lie in the range (0,1).
Note that, in deriving
Eq.(\ref{OGMs}), we have used charge symmetry to set 
${}^{\ell}G_M^u = {}^{\ell}G_M^d$. 
Since the chiral coefficients for the $d$ and $s$ loops in the RHS  
of Fig.~1 are identical, the main difference comes from the mass of the 
$K$ compared with that of the $\pi$.

With a little algebra $O_N$, and hence $G_M^s (\equiv {{}^{\ell}G_M^s})$, 
may be isolated
from Eqs.~(\ref{equalities}) and (\ref{OGMs}):
\begin{eqnarray}
G_M^s &=& \left ( {\,{}^{\ell}R_d^s \over 1 - \,{}^{\ell}R_d^s }
\right ) \left [ 2 p + n - {u^p \over u^{\Sigma}} \left ( \Sigma^+ -
\Sigma^- \right ) \right ] ,
\label{GMsSigma} \\
G_M^s &=& \left ( {\,{}^{\ell}R_d^s \over 
1 - \,{}^{\ell}R_d^s } \right ) \left [
p + 2n - {u^n \over u^{\Xi}} \left ( \Xi^0 - \Xi^- \right ) 
 \right ] .
\label{GMsXi}
\end{eqnarray}
Incorporating the experimentally measured baryon
moments \cite{PDG} (in nuclear magnetons, $\mu_N$), 
Eqs.~(\ref{GMsSigma}) and (\ref{GMsXi}) become:
\begin{eqnarray}
G_M^s &=& \left ( {\,{}^{\ell}R_d^s \over 1 - \,{}^{\ell}R_d^s } \right ) \left [
3.673 - {u^p \over u^{\Sigma}} \left ( 3.618 \right ) \right ] , 
\label{ok} \\
G_M^s &=& \left ( {\,{}^{\ell}R_d^s \over 
1 - \,{}^{\ell}R_d^s } \right ) \left [
-1.033 - {u^n \over u^{\Xi}} \left ( -0.599 \right ) \right ] ,
\label{great}
\end{eqnarray}
%
%
We stress that {\em these expressions for
$G_M^s$ are exact consequences of QCD, under the assumption of charge
symmetry}. 

\begin{figure}[tbp]
\begin{center}
{\includegraphics[height=7.8cm,angle=90]{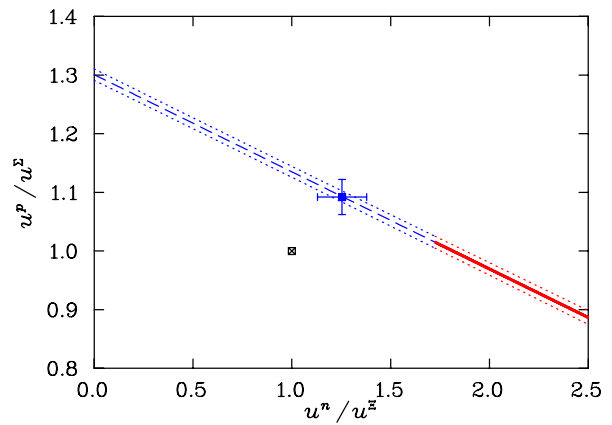}}
\end{center}
\vspace*{-0.5cm}
\caption{The constraint (dashed $G_M^s(0) < 0$, solid $G_M^s(0) > 0$)
on the ratios $u^p/u^{\Sigma}$ and $u^n/u^{\Xi}$ implied by charge
symmetry and experimental moments.  Experimental uncertainties are
indicated by the dotted bounds.  The assumption of environment independent
quark moments is indicated by the crossed square.  Our final result
(chiral corrected extrapolation of lattice results) is illustrated by the
filled square on the charge symmetry line.  }
\label{SelfCons}
\vspace{-0.5cm}
\end{figure}

Equating (\ref{ok}) and (\ref{great}) provides a linear relationship
between $u^p/u^{\Sigma}$ and $u^n/u^{\Xi}$ which must be satisfied
within QCD under the assumption of charge symmetry.  Figure
\ref{SelfCons} displays this relationship by the dashed and solid
line.  Since this line does not pass through the point $(1.0, 1.0)$,
corresponding to the simple quark model assumption of universality,
there must be an environment effect exceeding 12\% in both ratios or
approaching 20\% or more in at least one of the ratios.  Indeed, a
positive value for $G_M^s(0)$ would require an environment sensitivity
exceeding 70\% in the $u^n/u^{\Xi}$ ratio!

\begin{figure}[bp]
\vspace*{-0.3cm}
\begin{center}
  {\includegraphics[height=4cm,angle=90]{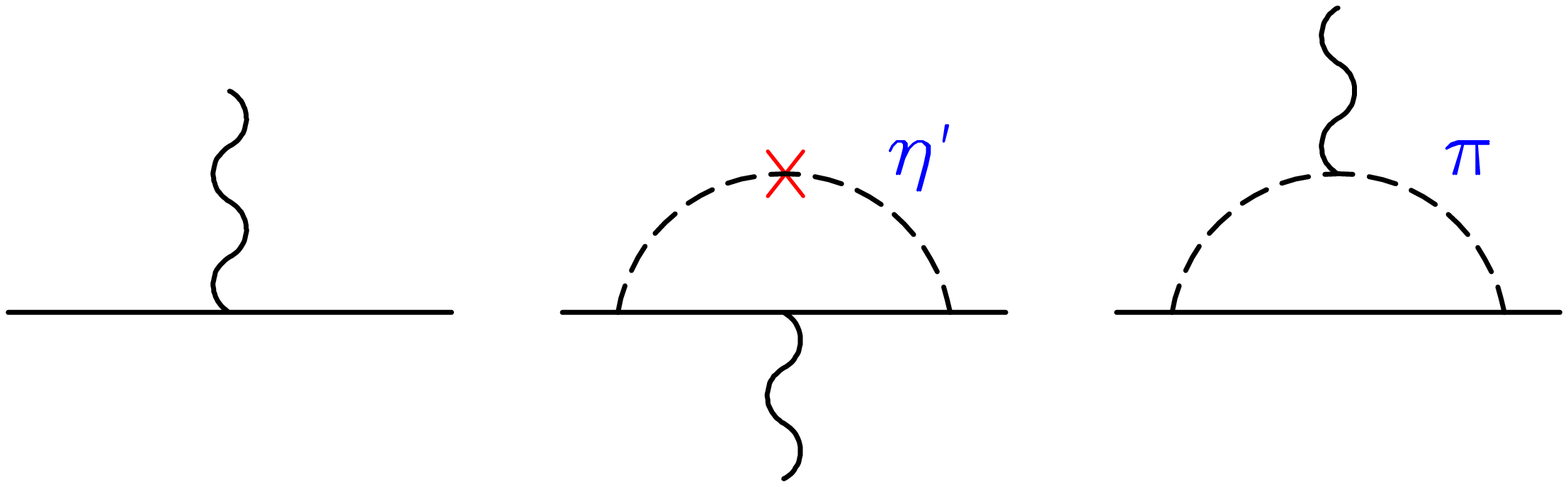}}

  {\includegraphics[height=4cm,angle=90]{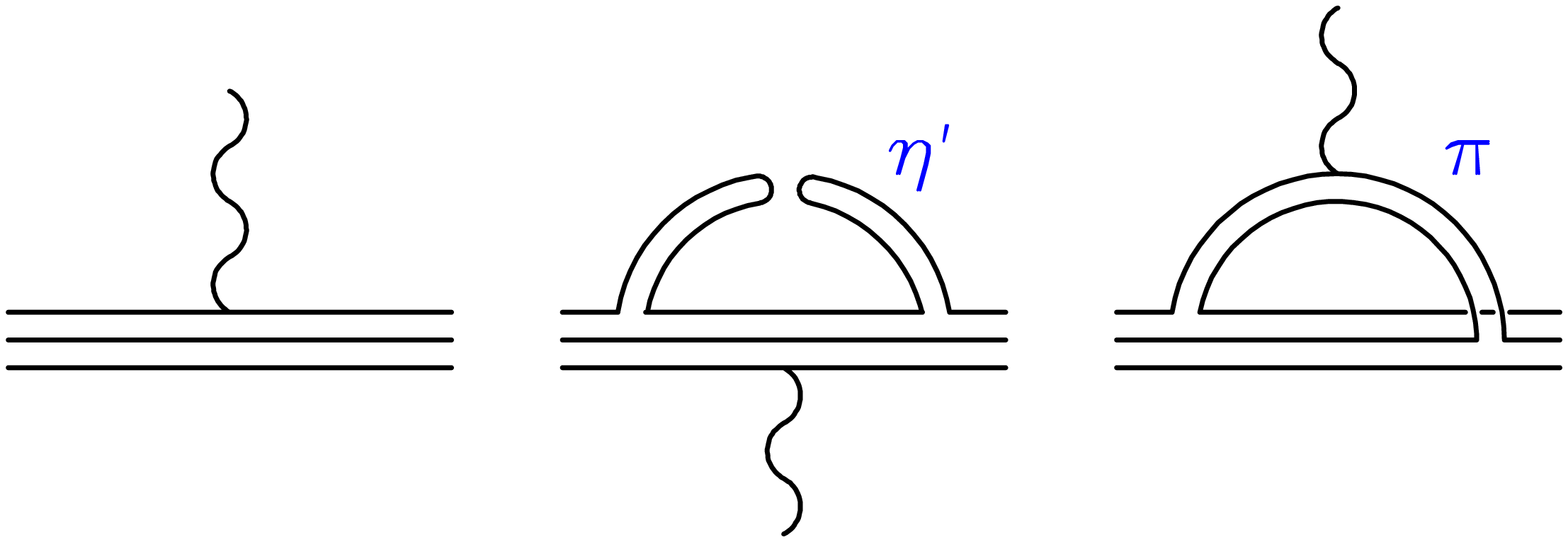}}
\end{center}
\vspace*{-0.5cm}
\caption{
Diagrams providing the leading contributions to the chiral expansion
of baryon magnetic moments (upper diagrams) and their associated quark
flows (lower diagrams) in QQCD.  
}
\label{QquarkFlow}
\vspace{-0.3cm}
\end{figure}

The numerical simulations of the
electromagnetic form factors presented here are carried out using the
Fat Link Irrelevant Clover (FLIC) fermion action
\cite{Zanotti:2001yb,Leinweber:2002bw} in which the irrelevant
operators, introduced to remove fermion doublers and lattice spacing
artifacts, are constructed with APE smeared links~\cite{ape}.
Perturbative renormalizations are small for smeared
links and the mean-field improved coefficients used here are
sufficient to remove ${\mathcal O}(a)$ errors from   
the lattice fermion action~\cite{Bilson-Thompson:2002jk}.

The ${\mathcal O}(a)$-improved conserved vector current
\cite{Martinelli:ny} is used.  Nonperturbative improvement is achieved
via the FLIC procedure, where the terms of the Noether current having
their origin in the irrelevant operators of the fermion action are
constructed with mean-field improved APE smeared links.  The
results presented here are obtained using established 
techniques~\cite{dblOctet} from a sample of 400 $20^3
\times 40$ mean-field ${\cal O}(a^2)$-improved Luscher-Weisz
\cite{Luscher:1984xn} 
gauge field configurations having a lattice spacing of 0.128 fm,
determined by the Sommer scale $r_0=0.50$ fm.

One of the major challenges in connecting lattice
calculations of hadronic properties with the physical world
\cite{dblPiCloud} is that 
currently accessible quark masses are 
much larger than the physical values.
Our present analysis has been
made possible by a significant breakthrough in the regularization of
the chiral loop contributions to hadron observables
\cite{Young:2002ib,Young:2004tb,Leinweber:2003dg}.  Through the process of
regulating loop integrals via a finite-range 
regulator~\cite{Young:2002ib,Donoghue:1998bs}, 
the chiral expansion is effectively re-summed to produce an
expansion with vastly improved convergence properties.
In particular, we extrapolate FLIC fermion calculations of
the valence quark contributions to baryon moments $(u^p,\ u^n,\ 
u^{\Sigma},\ u^{\Xi})$ to the physical mass regime.  We select the
dipole-vertex FRR with $\Lambda = 0.8$ GeV, which yields the best 
simultaneous description of both quenched and dynamical simulation
results~\cite{Young:2002cj}.

Separation of the valence and sea-quark-loop contributions to the
meson cloud of full QCD hadrons is a non-trivial task.  We use the
diagrammatic method for evaluating the quenched chiral coefficients of
leading nonanalytic terms in heavy-baryon quenched \chiPT\ (Q\chiPT ) 
\cite{Leinweber:2001jc,Leinweber:2002qb}.  
The valence contributions (key to this analysis)
are obtained by removing the direct-current coupling to sea-quark
loops from the total contributions.  Upon further removing ``indirect
sea-quark loop'' contributions, where a valence quark forms a meson
composed with a sea-quark loop, one obtains the ``quenched valence''
contributions -- the conventional view of the quenched approximation.

\begin{figure}[tbp]
\begin{center}
{\includegraphics[height=7.8cm,angle=90]{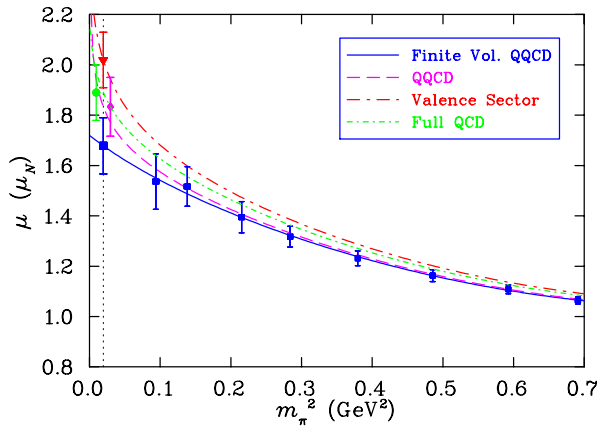}}
\end{center}
\vspace*{-0.5cm}
\caption{The contribution of a single $u$ quark (with unit
  charge) to the magnetic moment of the proton.  Lattice simulation
  results (square symbols for $m_\pi^2 > 0.05$ GeV) are extrapolated
  to the physical point (vertical dashed line) in finite-volume QQCD
  as well as infinite volume QQCD, valence and full QCD -- see text
  for details. 
  Extrapolated values 
  at the physical pion mass (vertical dashed
  line), are offset for clarity. }
\label{uProton}
\vspace{-0.2cm}
\end{figure}

\begin{figure}[tbp]
\begin{center}
{\includegraphics[height=4cm,angle=90]{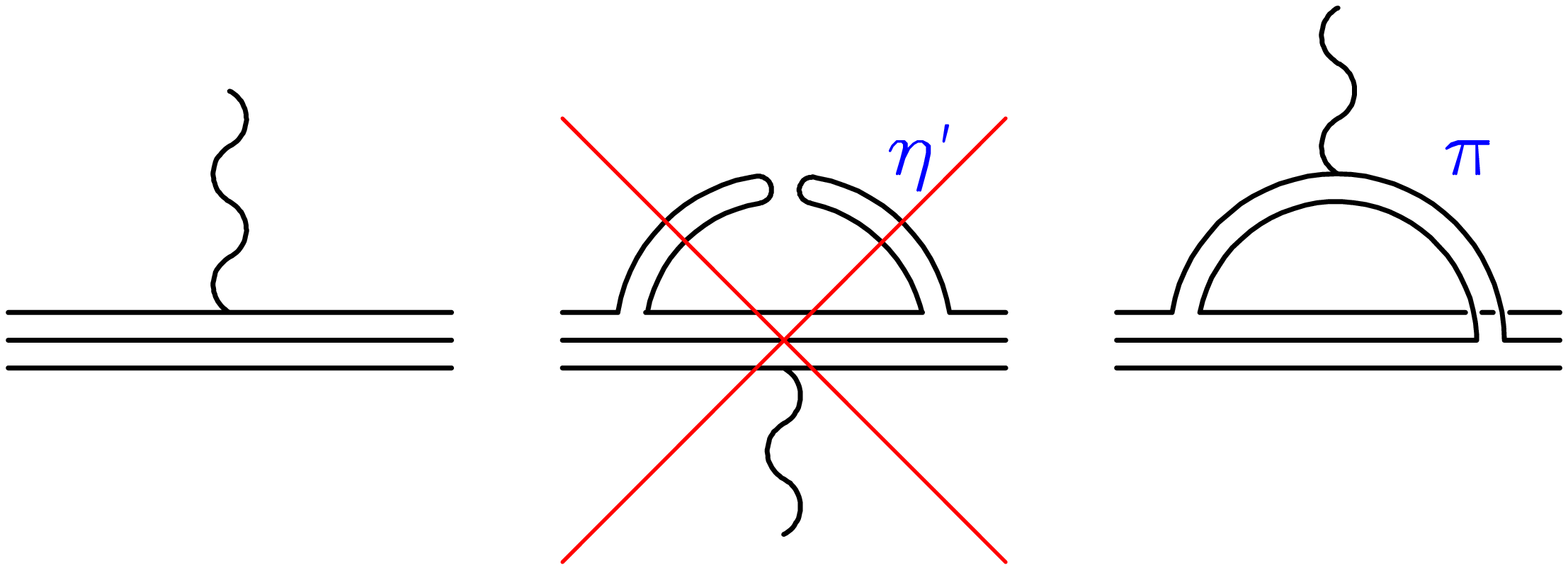}}

{\includegraphics[height=4cm,angle=90]{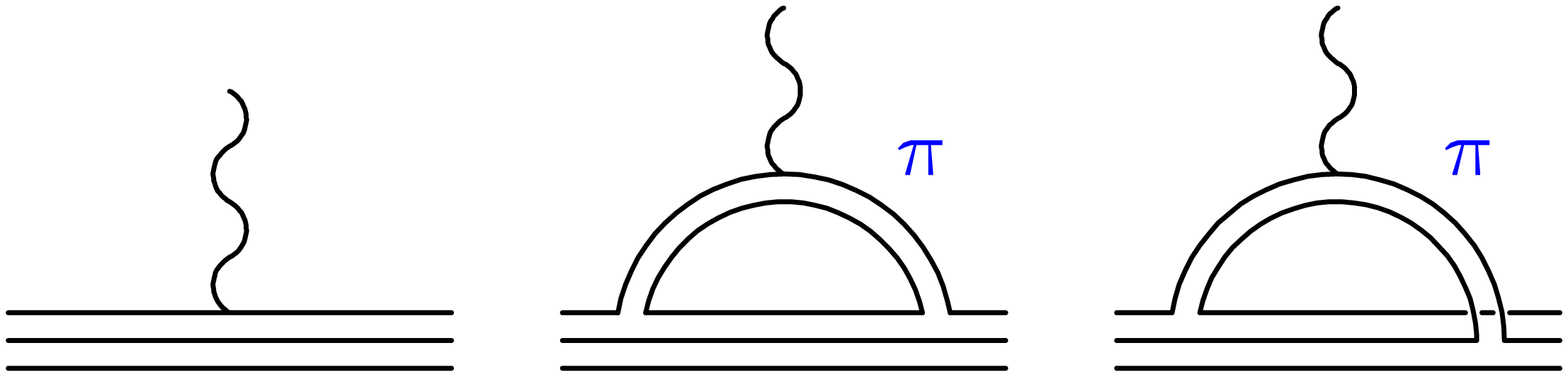}}
\end{center}
\vspace*{-0.5cm}
\caption{
Correcting Q\chiPT (upper) to the valence sector of full QCD (lower diagrams).
We remove quenched negative-metric $\eta'$ contributions and adjust
the chiral coefficients of $\pi$ and $K$ loops to
account for the coupling of a valence quark to the
photon in a meson made from a sea-quark loop. Coupling
to the anti-quark in the bottom-right diagram is also included in 
the valence contribution of full QCD.
}
\label{unquenching}
\vspace{-0.40cm}
\end{figure}

{}Figure~\ref{QquarkFlow} displays the diagrams providing the leading
contributions to the chiral expansion of baryon magnetic moments
(upper diagrams) and their associated quark flows in quenched QCD
(QQCD).  The associated chiral expansion for the proton magnetic
moment, $\mu_p$, has the form
\begin{eqnarray}
\mu_p &=& a_0^\Lambda 
            + \mu_p \, \chi_{\eta'} \, I_{\eta'}(m_\pi, \Lambda)
            + \chi_{\pi B} \, I_B(m_\pi, \Lambda) \nonumber \\
       &&   + \chi_{K B} \, I_B(m_K, \Lambda)
            + a_2^\Lambda \, m_\pi^2 
            + a_4^\Lambda \, m_\pi^4 \, .
\label{chiExp}
\end{eqnarray}
where the repeated index, $B$, sums over allowed baryon octet and decuplet
intermediate states.  Loop integrals denoted by $I$ are defined by
{\small
\begin{eqnarray}
&&\hspace*{-5mm}I_B(m, \Lambda)    = 
\nonumber\\&&\hspace*{-2mm}-\frac{2}{3\pi} \int  dk 
\frac{ \left (2 \sqrt{k^2+m^2} + \Delta_{BN} \right )\, k^4\, u^2(k,\Lambda)}{
\left (k^2+m^2 \right )^{3/2}\, \left (\sqrt{k^2+m^2} +
\Delta_{BN} \right )^2} \, ,\\
&&\hspace*{-5mm}I_{\eta'}(m_\pi, \Lambda) = - \int_0^\infty dk\,
\frac{k^4}{(k^2+m_\pi^2)^\frac{5}{2}}u^2(k,\Lambda) \, ,
\end{eqnarray}
}%
where $\Delta_{BN}$ is the relevant baryon mass-splitting and
$u(k,\Lambda)$ is the dipole-vertex regulator.  The coefficients,
$\chi$, denote the known model-independent coefficients of the LNA
term for $\pi$ and $K$ mesons \cite{Leinweber:2002qb,Savage:2001dy}.
We take $m_K^2 = m_K^{(0)\, 2} + \frac{1}{2} \, m_\pi^2$, and use the
physical values to define $ m_K^{(0)}$.  The $m_\pi^4$ term in
Eq.~(\ref{chiExp}) allows for some curvature associated with the Dirac
moment of the baryon, which should go as $1/m_\pi^2$ for moderately
large quark masses.

{}Figure \ref{uProton} illustrates a fit of FRR Q\chiPT to the FLIC
fermion lattice results (solid curve), where only the
discrete momenta allowed in the finite volume of the lattice are
summed in performing the loop integral.  The long-dashed curve that
also runs through the lattice results corresponds to replacing the 
discrete momentum sum by the infinite-volume, continuous
momentum integral.  For all but the lightest quark mass, finite volume
effects are negligible.

The coefficients of the residual expansion, $a_0^\Lambda,\
a_2^\Lambda,\ a_4^\Lambda$, show excellent signs of convergence.  For
example, the fit to $u^\Sigma$ yields values 1.48(7),
$-$0.90(23), and 0.42(19) in appropriate powers of GeV, respectively.
Incorporating baryon mass splittings into the kaon loop contributions
is essential -- e.g., the contribution of $\Sigma \to N K$ is almost 
doubled when the $\Sigma - N$ mass splitting is included.

\begin{figure}[tbp]
\begin{center}
{\includegraphics[height=7.8cm,angle=90]{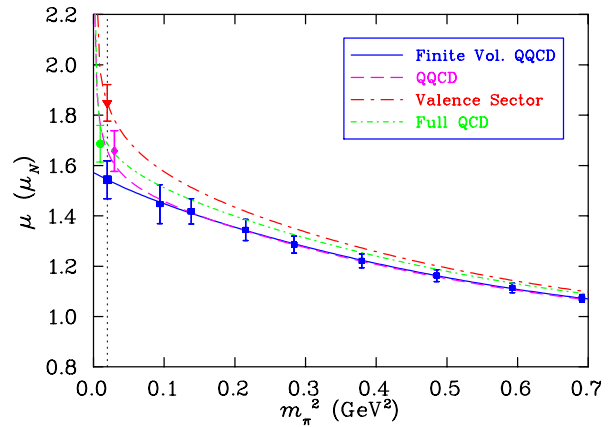}}
\end{center}
\vspace*{-0.5cm}
\caption{
The contribution of a single $u$ quark (with unit
charge) to the magnetic moment of the $\Sigma^+$.  Curves and symbols are
as for Fig.~\protect\ref{uProton}.
}
\label{uSigma}
\vspace{-0.5cm}
\end{figure}

{}Figure~\ref{unquenching} illustrates the
considerations in correcting the quenched $u$-quark contribution to
yield the valence $u$-quark contribution in full QCD.  The removal of
quenched $\eta'$ contributions and the appropriate adjustment of $\pi$
and $K$ loop
coefficients~\cite{Leinweber:2001jc,Leinweber:2002qb,Savage:2001dy} 
provides the dot-dash
curve of Fig.~\ref{uProton}.  This is our best estimate of the valence
$u$-quark contribution (connected insertion) to the
proton magnetic moment of full QCD.  
Finally, the disconnected insertion of the current is included to
estimate the total contribution of the $u$-quark sector to the proton
magnetic moment
\cite{Leinweber:2001jc,Leinweber:2002qb,Savage:2001dy}
(fine dash-dot curve in Fig.~\ref{uProton}).  Figure~\ref{uSigma}
displays similar results for the $\Sigma^+$.

%
%
%

From these chiral extrapolations, we estimate the ratios of the
valence (connected) $u$-quark contributions, ${u^p}/{u^\Sigma}$ and
${u^n}/{u^\Xi}$. 
The final results
\begin{equation}
\frac{u^p}{u^\Sigma} = 1.092\pm 0.030
\quad
\mbox{and}
\quad
\frac{u^n}{u^\Xi} = 1.254\pm 0.124
\label{uRatios}
\end{equation}
are plotted in Fig.~\ref{SelfCons}.  
The precision of these results follows from the use of correlated
ratios of moments which act to reduce uncertainties associated with
the lattice spacing, the regulator mass and statistical fluctuations
\cite{systematic}. 
This result leaves no doubt that $G_M^s$ is negative.  The fact that
this point lies exactly on the constraint curve is highly nontrivial,
and provides a robust check of the validity of the analysis techniques
presented here.

\begin{figure}[tbp]
\begin{center}
{\includegraphics[height=\hsize,angle=90]{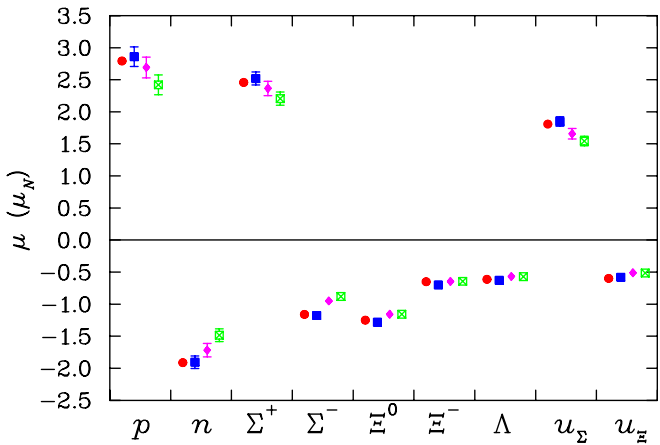}}
\end{center}
\vspace*{-0.5cm}
\caption{ The one standard deviation agreement between the \FRRchiPT
corrected lattice simulation results ($\blacksquare$) and the
experimentally measured baryon magnetic moments ($\bullet$).  Quenched
($\blacklozenge$) and finite-volume quenched ($\boxtimes$) results are
also illustrated.  }
\label{compExpt}
\vspace{-0.5cm}
\end{figure}

As a further check, in Fig.~\ref{compExpt} we compare the lattice QCD
predictions of the baryon magnetic moments constructed from
chirally-corrected extrapolations of the individual quark sectors.
The results display an unprecedented level of agreement with
experiment.  We note that the experimental constraints on $u^\Sigma$
and $u^\Xi$ emphasized by Wong~\cite{Wong:2001nm} are both satisfied
precisely.
%
%
%

While $G_M^s$ is most certainly negative, it remains to set the
magnitude.  This requires an estimate of the strange to light
sea-quark loop contributions, ${}^{\ell}R_d^s$.  Earlier estimates of
$^{\ell}R_d^s$ were based on the constituent quark model.  A more
reliable approach is to estimate the loops using the same successful
model invoked to correct the quenched results to full QCD
\cite{Young:2002cj,Young:2004tb} as illustrated in Fig.~\ref{compExpt}.
Allowing the dipole mass parameter to vary between 0.6 and 1.0 GeV
provides ${}^{\ell}R_d^s = {G_M^s}/{G_M^d}=0.139 \pm 0.042$.
%
%
A complete analysis of the errors associated with the determination of
$G_M^s$ using Eqs.~(\ref{GMsSigma}), (\ref{GMsXi}) and (\ref{uRatios})
is reported in Ref.~\cite{systematic}.  The uncertainty is dominated
by the statistical errors included in Eq.~(\ref{uRatios}) and the
uncertainty just noted for ${}^{\ell}R_d^s$.  The final result for the
strangeness magnetic moment of the nucleon is
\begin{equation} 
{G_M^s} = -0.046 \pm 0.019\ \mu_N \, .
\label{GMs}
\end{equation} 
This precise value sets a tremendous challenge for the next generation
of parity violation experiments.

\vspace{6pt}

We thank the Australian Partnership for Advanced Computing (APAC) for
generous grants of supercomputer time which have enabled this project.
This work is supported by the Australian Research Council
and by DOE contract DE-AC05-84ER40150, under which SURA operates
Jefferson Laboratory.


\vspace{-12pt}

\begin{thebibliography}{30}
%
\bibitem{Aniol:2000at}
\vspace*{-12pt}
K.~A.~Aniol {\it et al.}  [HAPPEX Collaboration],
Phys.\ Lett.\ B {\bf 509}, 211 (2001)
[nucl-ex/0006002].
%
\bibitem{Spayde:2003nr}
D.~T.~Spayde {\it et al.}  [SAMPLE Collaboration],
Phys.\ Lett.\ B {\bf 583}, 79 (2004)
[arXiv:nucl-ex/0312016].
%
\bibitem{Mathur:2000cf}
N.~Mathur and S.~J.~Dong, 
Nucl.\ Phys.\ Proc.\ Suppl.\  {\bf 94}, 311 (2001)
[hep-lat/0011015];
S.~J.~Dong, {\it et al.},
Phys.\ Rev.\ {\bf D58}, 074504 (1998)
[hep-ph/9712483].

\bibitem{Lewis:2002ix}
R.~Lewis, W.~Wilcox and R.~M.~Woloshyn,
Phys.\ Rev.\ D {\bf 67}, 013003 (2003)
[hep-ph/0210064].

\bibitem{Zanotti:2001yb}
J.~M.~Zanotti {\it et al.}  [CSSM Lattice Collaboration],
Phys.\ Rev.\ D {\bf 65}, 074507 (2002)
[hep-lat/0110216].

\bibitem{Leinweber:2002bw}
D.~B.~Leinweber {\it et al.},
Eur.\ Phys.\ J. A {\bf 18}, 247 (2003)
[nucl-th/0211014].

\bibitem{FLICscaling}
J.~M.~Zanotti, B.~Lasscock, D.~B.~Leinweber, and A.~G.~Williams,
hep-lat/0405015.

\bibitem{FLIClqm}
J.~M.~Zanotti, S.~Boinepalli, W.~Kamleh, D.~B.~Leinweber, and A.~G.~Williams,
hep-lat/0405026.

\bibitem{Young:2002ib}
R.~D.~Young, D.~B.~Leinweber and A.~W.~Thomas,
Prog.\ Part.\ Nucl.\ Phys.\  {\bf 50}, 399 (2003)
[hep-lat/0212031].

\bibitem{Young:2004tb}
R.~D.~Young, D.~B.~Leinweber and A.~W.~Thomas,
Phys.\ Rev.\ D {\bf 71}, 014001 (2005)
[arXiv:hep-lat/0406001].

\bibitem{ChargeSymm}
G.A.~Miller {\it et al.},
Phys.\ Rept.\ {\bf 194}, 1 (1990).
%
\bibitem{Leinweber:1996ie}
D.B.~Leinweber,
Phys.\ Rev.\ {\bf D53}, 5115 (1996)
[hep-ph/9512319].
%
\bibitem{Leinweber:1999nf}
D.~B.~Leinweber and A.~W.~Thomas,
Phys.\ Rev.\ D {\bf 62}, 074505 (2000)
[hep-lat/9912052].
%
\bibitem{PDG}
Particle Data Group, Phys.\ Rev.\ {\bf D66}, 010001 (2002).
%
\bibitem{ape}
M.~Falcioni {\it et al.},
Nucl.\ Phys.\ {\bf B251}, 624 (1985).
%
\bibitem{Bilson-Thompson:2002jk}
S.~O.~Bilson-Thompson {\it et al.},
Ann.\ Phys.\  {\bf 304}, 1 (2003)
[hep-lat/0203008].
%
\bibitem{Martinelli:ny}
G.~Martinelli {\it et al.},
Nucl.\ Phys.\ B {\bf 358}, 212 (1991)
%
\bibitem{dblOctet}
D.~B. Leinweber {\it et al.},
Phys.\ Rev.\ {\bf D43}, 1659 (1991).
%
\bibitem{Luscher:1984xn}
M.~Luscher and P.~Weisz,
Commun.\ Math.\ Phys.\  {\bf 97}, 59 (1985)
[ibid.\  {\bf 98}, 433 (1985)].
%
\bibitem{dblPiCloud}
T.~D. Cohen and D.~B. Leinweber, 
Comments Nucl.\ Part.\ Phys.\ {\bf 21}, 137   (1993)
[hep-ph/9212225];
A.W.~Thomas,
Austral.\ J.\ Phys.\ {\bf 44}, 173 (1991).
%
%
\bibitem{Leinweber:2003dg}
D.~B.~Leinweber, A.~W.~Thomas and R.~D.~Young,
Phys.\ Rev.\ Lett.\  {\bf 92}, 242002 (2004)
[hep-lat/0302020].
%
\bibitem{Donoghue:1998bs}
J.~F. Donoghue, {\it et al.},
Phys.\ Rev.\ {\bf D59}, 036002 (1999).
%
%
%
%
%
\bibitem{Young:2002cj}
R.~D.~Young {\it et al.},
Phys.\ Rev.\ D {\bf 66}, 094507 (2002)
[hep-lat/0205017].

\bibitem{Leinweber:2001jc}
D.~B.~Leinweber,
Nucl.\ Phys.\ Proc.\ Suppl.\  {\bf 109A}, 45 (2002)
[hep-lat/0112021].

\bibitem{Leinweber:2002qb}
D.~B.~Leinweber,
Phys.\ Rev.\ D {\bf 69}, 014005 (2004)
[hep-lat/0211017].

\bibitem{Savage:2001dy}
M.~J.~Savage,
Nucl.\ Phys.\ A {\bf 700}, 359 (2002)
[nucl-th/0107038].

\bibitem{systematic}
D.~B.~Leinweber, {\it et al.}, hep-lat/0502004.

\bibitem{Wong:2001nm}
C.~W.~Wong,
hep-lat/0103021.

\end{thebibliography}
\end{document}